\documentclass{appolb}
\usepackage[dvips]{graphicx}
\usepackage{amsmath}
\newcommand {\dperp} {d_{T}}
\newcommand {\dperpc} {d_{T,C}}
\newcommand {\dperppi} {d_{T,\pi}}

\begin{document}
\title{Colour dipoles and deeply virtual Compton scattering}%

\author{Ruben Sandapen
\address{Department of Physics and Astronomy, \\
University of Manchester, M13 9PL, UK\\\textit{ruben@theory.ph.man.ac.uk}}}
\maketitle

\begin{abstract}
\noindent
I report on an  analysis of  Deeply Virtual Compton Scattering (DVCS) within the dipole model, done in collaboration with M.~McDermott and G.~Shaw~\cite{mss}.  The two models considered here are distinct in their structure and implications. They both agree with the available cross-section data on DVCS from HERA~\cite{h1data}. Predictions for various asymmetries are also given.  
\end{abstract}
\medskip
Why DVCS ? On the theory side, we have an explicitly proven factorisation theorem~\cite{jcaf}, valid for \textit{asymptotic} $Q^2$, which expresses the amplitude as a convolution in perturbatively calculable coefficients with generalised parton distributions\footnote{Leading order and next to leading order calculations have since been done~\cite{fm,afmm}.}.  A dipole analysis\footnote{Donnachie and Dosch provided a dipole analysis in~\cite{dd}.}of DVCS is complementary to this formal QCD analysis: going beyond leading twist,  we can, in principle, establish quantitatively a regime in $Q^2$ for which the formal pQCD approach is valid.  Furthermore,  at the lepton level, the interference of DVCS with the purely real Bethe-Heitler (BH) process  offers a unique opportunity to isolate the real and imaginary DVCS amplitude via various azimuthal angle asymmetries~\cite{belitsky}.\\    

The dipole model  is valid at low $x$, when a factorisation of time-scales allows us to express the  forward  diffractive amplitude as a convolution in the \textit{photon wave functions} and the \textit{dipole cross-section} as shown in~(\ref{convolution}). 

\begin{equation}
\mathcal{A}(s,t=0)= s\int dz~d^2\dperp~\Psi^{*}_{\gamma^{*}}(\dperp,z,Q^2)~\sigma_{d}~\Psi_{\gamma}(\dperp,z,0)
\label{convolution}
\end{equation}
\noindent
The conserved quantities during the interaction are the dipole size $\dperp$ and the longitudinal momentum fraction $z$  carried by the quark. Since the final state photon is real,  the final  photon wave function is evaluated at $Q^2=0$.   For small dipole size $\dperp$, the photon wave function can be calculated perturbatively using the usual $-i\gamma^\mu$ QED vertex~(see~\textit{e.g}~\cite{gieseke}). The challenge is to model the dipole cross-section\footnote{A number of authors have proposed different models, see \textit{e.g}~\cite{amirim}~for an overview.}, $\sigma_{d}$, which encodes  all the dynamics (perturbative and non-perturbative) of the dipole-proton interaction. Colour transparency dictates $\sigma_{d}$ to  vanish as $r^2$ as $r^2\rightarrow0$.  On purely geometrical grounds, we expect a monotonic increase of $\sigma_d$  with $\dperp$ and $\sigma_d$ to  become \textit{hadron-like} at large $\dperp$. We assume no flavour and $z$ dependence. As for the energy dependence, dipole models fall into two main classes : either direct dependence on $W$ or via $x$.  It is an ongoing issue whether saturation effects should be incorporated in dipole models. \textit{One of the models presented here does include saturation effects while the other does not.}\\ 

The FKS (Forshaw, Kerley, Shaw) dipole cross-section~\cite{fks1} is a sum of a soft and a hard term, each with Regge-like energy dependence. 
\begin{equation}
\hat {\sigma}  (W^2,\dperp)= a_{0}^{S}~P^{\mbox{s}}(a_n^s,\dperp)(\dperp^{2} W^2)^{\lambda_{S}} + P^{\mbox{h}}(a_n^h,\dperp)~\exp (-\nu_{H} \dperp)~(\dperp^{2} W^2)^{\lambda_{H}}
\end{equation}
$P^{\mbox{s}}(a_n^s,\dperp)$ and $P^{\mbox{h}}(a_n^h,\dperp)$ are polynomials in $ \dperp$.  A distinctive feature of the FKS model is that the authors modify the photon wave function at large $\dperp$ using a shifted Gaussian, $f(\dperp)$ :~$|\psi(z, \dperp, Q^2)|^2 \rightarrow |\psi(z,\dperp, Q^2)|^2.f(\dperp)$~\cite{fks1}.  All the free parameters in the FKS have been successfully  fitted to $F_{2}$ and real photoabsorption data~\cite{fks1} and the model has been used to predict $F_{2}^D$~\cite{fks2}. In its present form, the FKS model does not include saturation effects.  We use this fitted dipole cross-section to make a no-free parameter predictions for DVCS. \\

The MFGS (McDermott, Frankfurt, Guzey and  Strikman) model~\cite{mfgs}  makes direct contact with pQCD, as for small $\dperp$ ($\dperp < \dperpc$), it is directly related to the gluon distribution in the proton, through the well-known equation~\cite{derivation}
\begin{equation}
\hat{\sigma}_{\mbox{{\tiny pQCD}}}(x,\dperp) =\frac{\pi^{2} \dperp^{2}}{3}~\alpha_{s}(\bar{Q}^{2})~xg(x_{g},\bar{Q}^2)\hspace{1cm};\hspace{1cm}\bar{Q}^{2} = \frac{\lambda}{\dperp^{2}}
\label{pqcddipole}
\end{equation}
In an attempt to go beyond leading log, $\dperp$-dependence is included in the scales $\bar{Q}$ and $x_{g}$~\cite{mfgs}. For large dipole size, $\dperp > \dperppi$, where $\dperppi$ is the pion size, $\sigma_d$ is matched onto the pion-proton cross-section. The dipole cross-section is linearly interpolated between $\dperpc$ and $\dperppi$. What is the appropriate value for $\dperpc$? At moderate $x$ , $\dperpc = 0.246~\mbox{fm}$, corresponding to  $Q = Q_0 = 1.6 ~\mbox{GeV}$. However, for sufficiently low $x$, the strong rise in the gluon distribution makes the dipole cross-section for small $\dperp$ exceed that for large $\dperp$.  To prevent this, $\dperpc$ is made to shift to increasingly small values as $x$ decreases. In this way, saturation effects are included. \textit{For $\lambda =4 $, this correction is not important for the HERA region but does become important above it}. For an exclusive process such as DVCS, it is necessary to use the generalised (or skewed) gluon distribution in ~(\ref{pqcddipole}), which has an additional dependence (as compared to the ordinary gluon distribution) on the skewedness parameter $\delta = x$. For our calculation,  we adapted the (leading order) skewed evolution package of Freund and Guzey~\cite{fg}, with CTEQ4L gluon distributions as input.\\

\begin{figure}[htbp]
   \includegraphics[width=5.5cm,height=3.5cm]{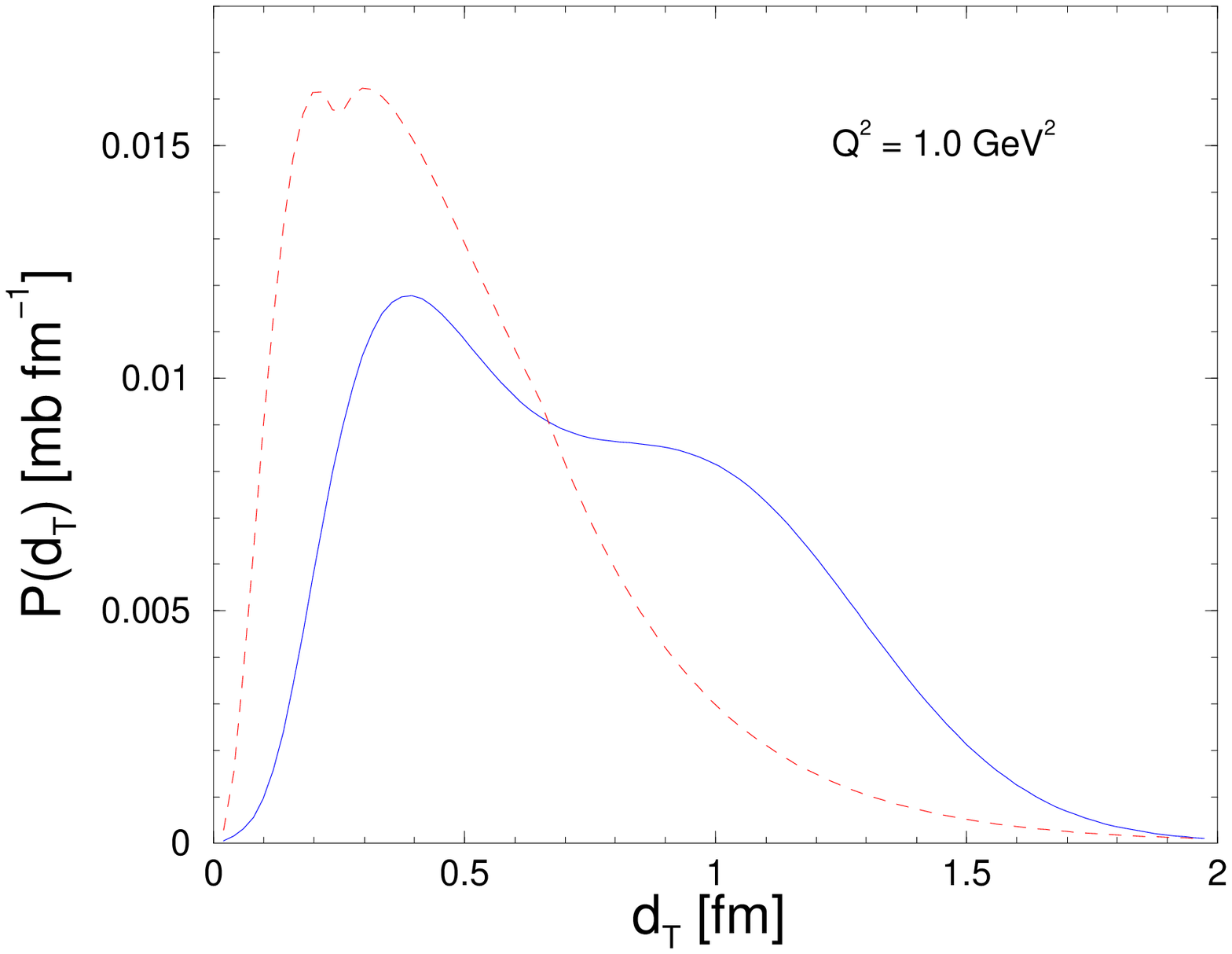}
  \hspace{1cm} \includegraphics[width=5.5cm,height=3.5cm]{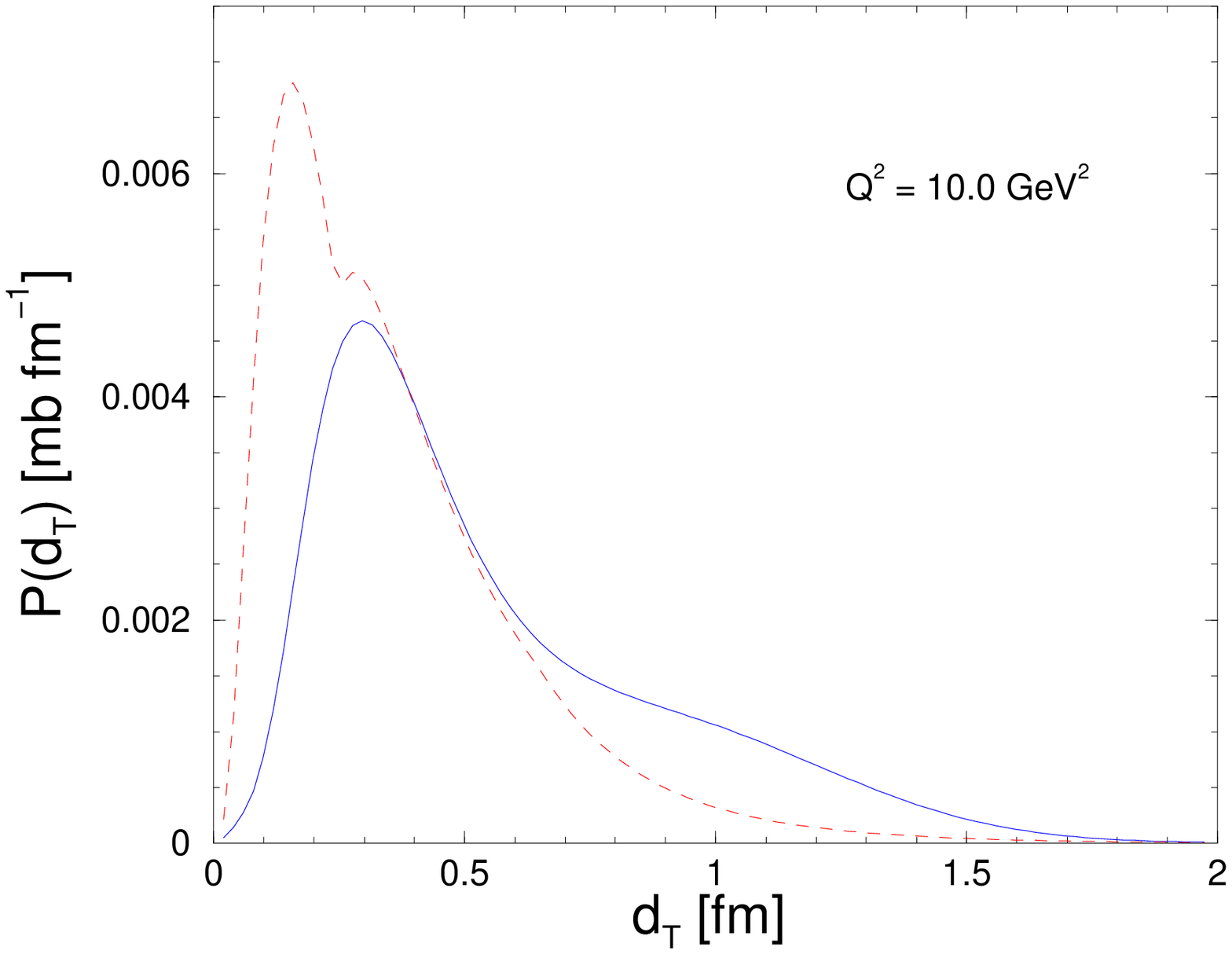}
  \caption{Profiles  in transverse dipole size for different $Q^2$ values and $W = 75$~GeV, 
FKS (solid line) and MFGS (dashed line).}
    \label{profiles.eps}
\end{figure}

\begin{figure}[htbp]
   \includegraphics[width=5.5cm,height=3.5cm]{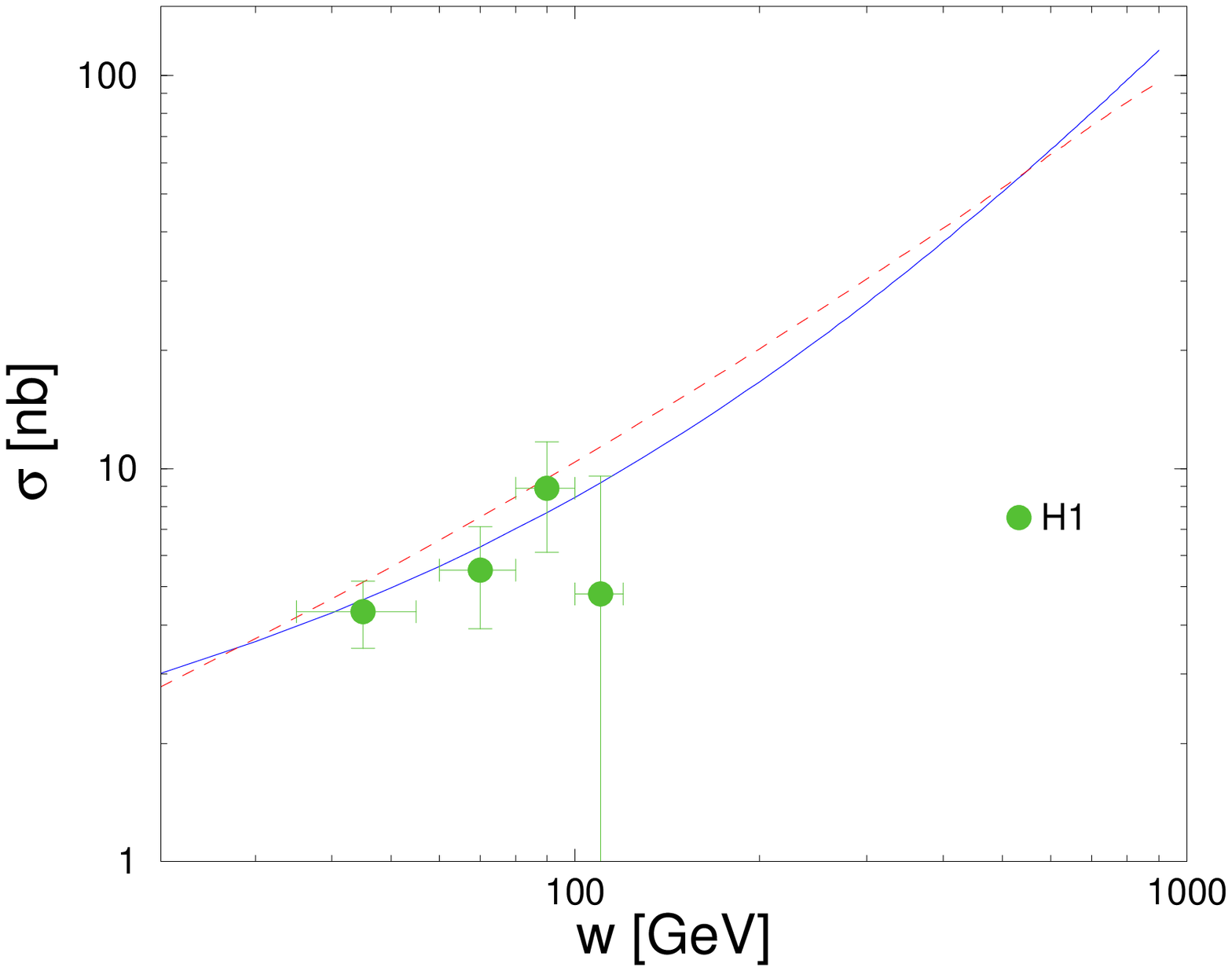}
  \hspace{1cm} \includegraphics[width=5.5cm,height=3.5cm]{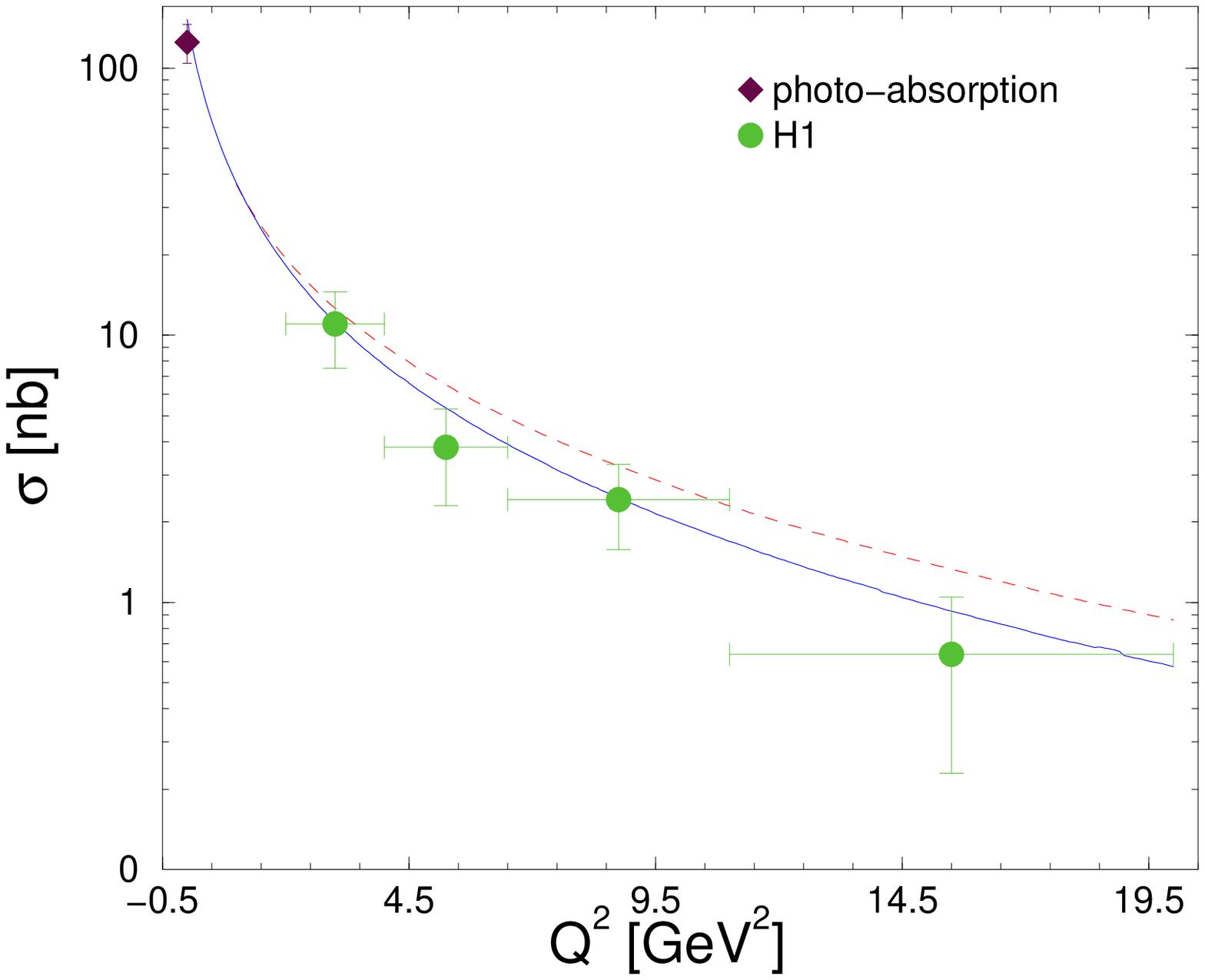}
  \caption{The energy (at $Q^2=4.5~\mbox{GeV}^2$) dependence and $Q^2$-dependence (at $W = 75$~GeV) of the photon level DVCS cross-section, FKS (solid line) and MFGS (dashed line).}
    \label{totsig.eps}
\end{figure}

\begin{figure}[htbp]
   \includegraphics[width=5.5cm,height=3.5cm]{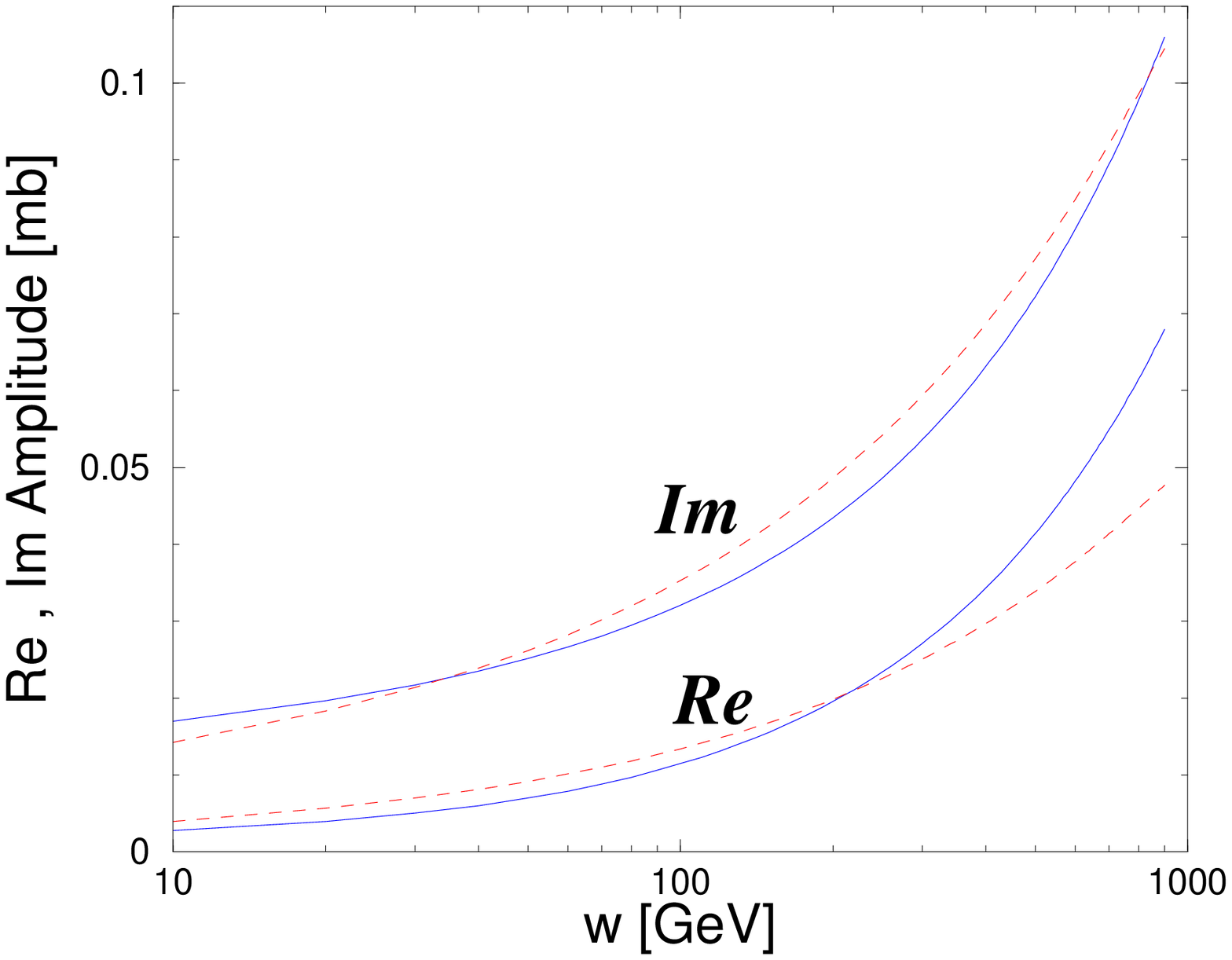}\hspace{1cm} \includegraphics[width=6.0cm,height=3.5cm]{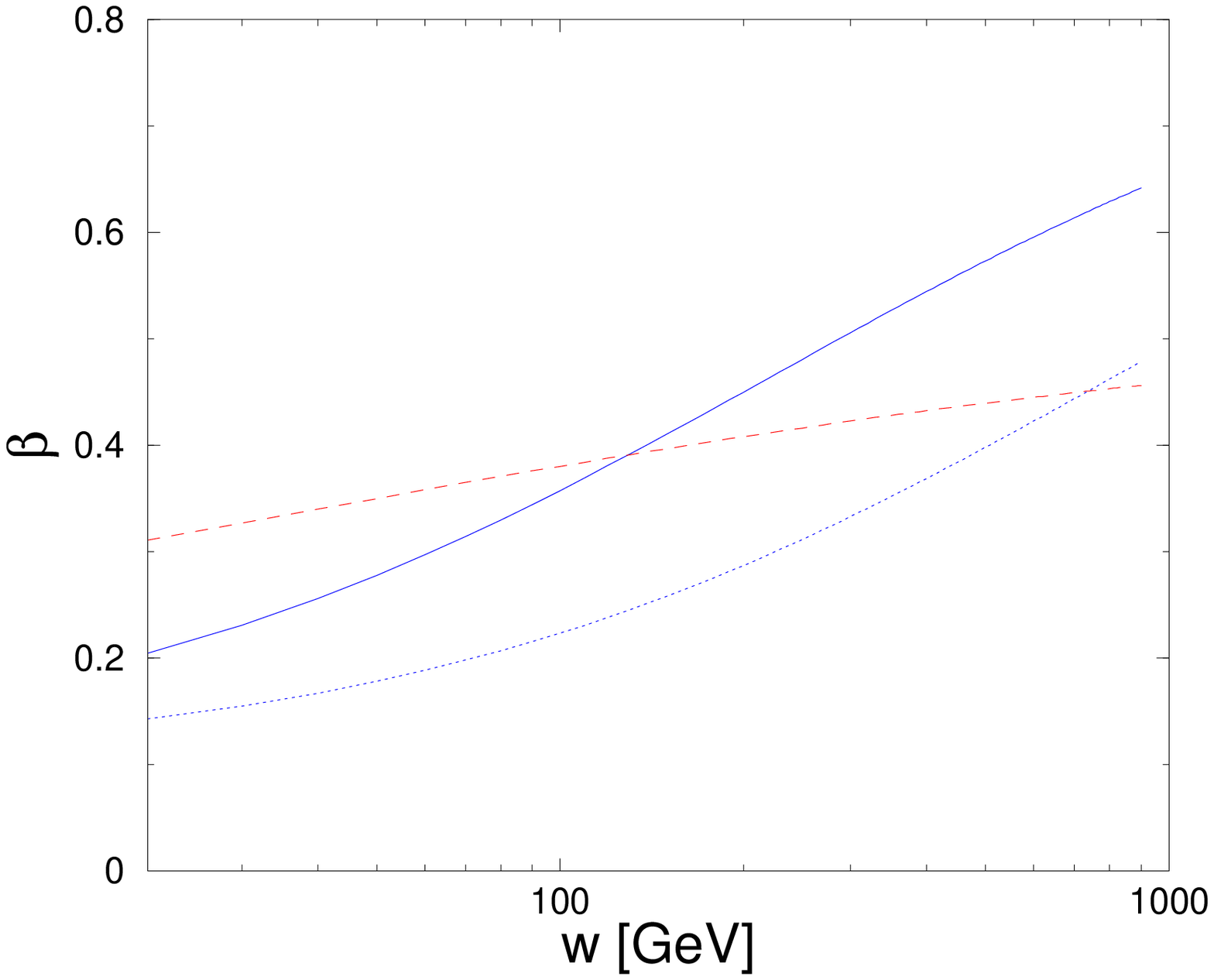}
   \caption{The energy dependence (at $Q^2=4.5~\mbox{GeV}^2$) of the real and imaginary parts of the DVCS  amplitude, FKS (solid line) and MFGS (dashed line). $\beta$ is the real to imaginary parts ratio. Dotted line for FKS model, at $Q^2 = 0$.}
    \label{realamp.eps}
\end{figure}

To see the relative contribution of different dipole sizes to the amplitude, we integrate out the angular and $z$ dependence in Eq.~\ref{convolution}. The results are shown in Fig \ref{profiles.eps}.
The FKS model has a larger contribution from large dipoles than the MFGS model although the forward amplitudes obtained by integrating over $\dperp$ are similar.  We assume the usual exponential $t$ dependence of the differential cross-section  and use a value of $7~\mbox{GeV}^{-2}$\footnote{This is the value used by H1 in their analysis.} for the slope parameter $B$ to calculate the photon-level total cross-section, $\sigma(\gamma^* p\rightarrow\gamma p)$. The theoretical predictions, which include the small contribution of the real part of the amplitude, are compared to H1 data points in Fig~\ref{totsig.eps}.  We reconstruct the real amplitude using analyticity. The FKS amplitude being a sum of two Regge terms, the real part is easily computed using the signature factors. As for the MFGS amplitude, we have to do  a  two power fit to the imaginary amplitude first.  The FKS real amplitude shows a steeper energy dependence at very high energies as can be clearly seen in Fig~\ref{realamp.eps}. \\

\begin{figure}[htbp]
   \includegraphics[width=5.5cm,height=3.5cm]{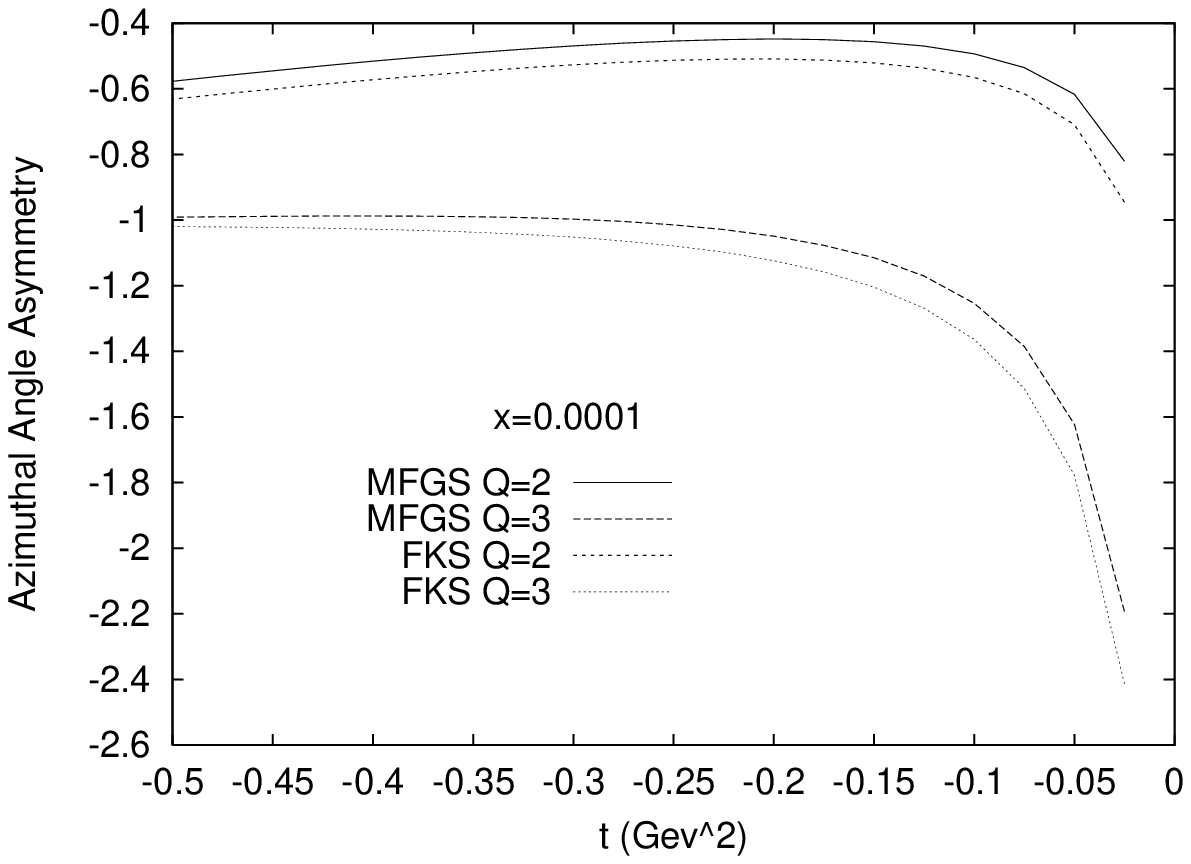}
  \hspace{1cm}\includegraphics[width=5.5cm,height=3.5cm]{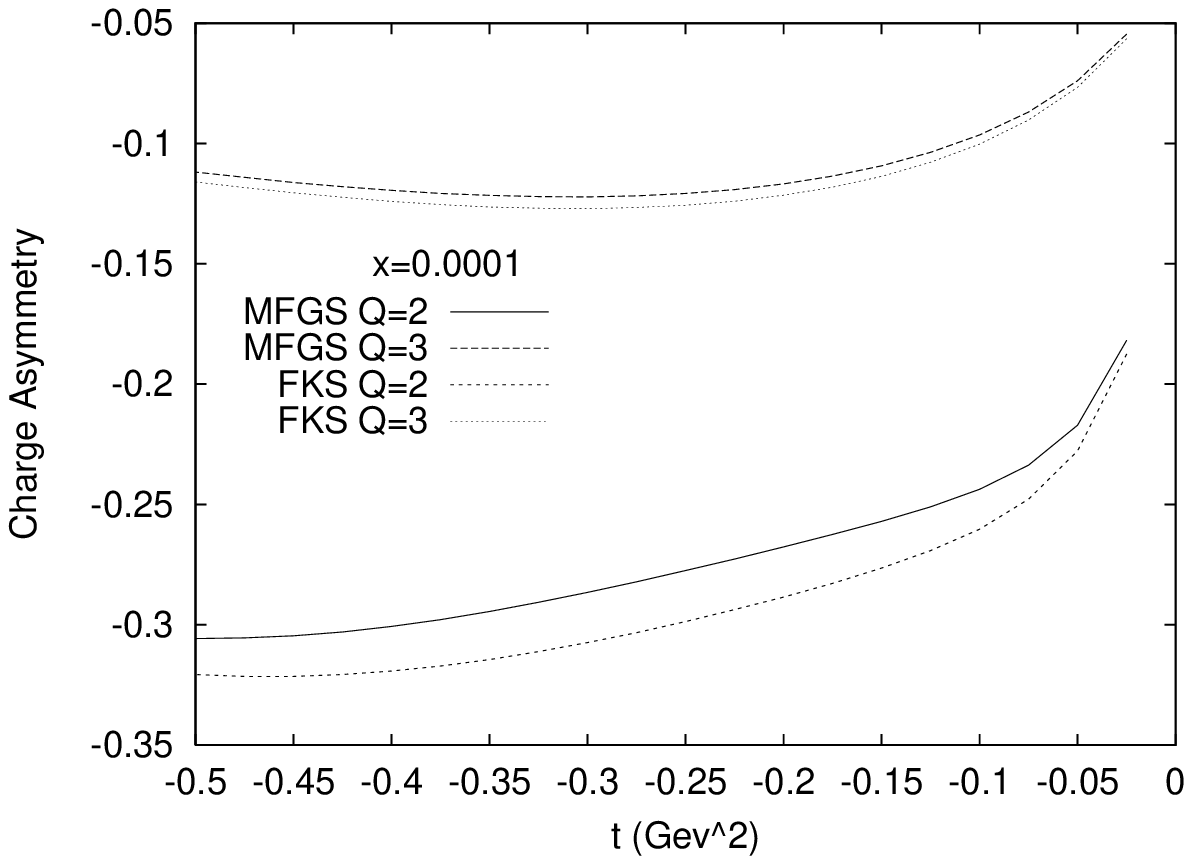}
  \caption{Asymmetries for fixed $x=10^{-4}$ at two values of $Q = 2,3~\mbox{GeV}^2$, accessible in the HERA kinematic range.}
    \label{asymmetries.eps}
\end{figure}

As mentioned in the introduction, we can isolate the real and imaginary parts of the DVCS amplitude via various azimuthal angle asymmetries, defined and discussed in ~\cite{mss}. I simply highlight here that in the special frame of reference~\cite{mss,belitsky} chosen for our calculation, the pure BH cross-section  has a residual $\phi$ dependence. It is necessary to subtract off the BH contribution in defining the azimuthal angle asymmetry (AAA) (see eqn. 37 of ~\cite{mss}) so that AAA becomes directly proportional to the real DVCS amplitude in the joint limit of low $x$ and high $Q^2$.  In this limit, the charge asymmetry (eqn. (39) in~\cite{mss}) is also directly proportional to the real DVCS amplitude, while the single spin asymmetry (SSA)(eqn. (38) in~\cite{mss}) is propotional to the imaginary DVCS amplitude. Our predictions for AAA and CA are shown in Fig~\ref{asymmetries.eps}\footnote{M.~McDermott adapted a code from~\cite{fm} to produce Fig~\ref{asymmetries.eps}.}.\\  

We have used two different dipole models to make predictions for DVCS. There is good agreement with H1 data for both models, even beyond the HERA region, despite only one of them including saturation effects.  A more pronounced difference is found for the real amplitude, at very high energies.  Experimental measurements of these asymmetries will allow us to test our predictions.\\ 

It is a pleasure to thank the organisers for this very interesting workshop. I am also happy to thank M.~McDermott, J.~Forshaw and G.~Shaw for helpful discussions and the University of Manchester for financial support.

\end{document}